\documentclass[manuscript,screen, authorversion,nonacm]{acmart}

\usepackage{tcolorbox}
\tcbuselibrary{breakable}
\usepackage{tikz}
\usetikzlibrary{positioning}
\usetikzlibrary{matrix,calc}
\usetikzlibrary{backgrounds}
\usepackage{multirow}
\usepackage{multicol}

\usepackage{colortbl}
\usepackage{enumitem}
\usepackage{ifthen}
\usepackage{tabularx}
\usepackage{tabularray}
\usepackage{rotating}
\usepackage{makecell}
\usepackage{hhline}
\usepackage{nicematrix}
\usepackage{booktabs}
\usepackage{array}
\usepackage{adjustbox}
\usepackage{lipsum}
\usepackage{subfig}
\usepackage{ragged2e} % for \raggedright command
\usepackage{url}

\newcommand{\PreserveBackslash}[1]{\let\temp=\\#1\let\\=\temp}
\newcolumntype{C}[1]{>{\PreserveBackslash\centering}p{#1}}
\newcolumntype{R}[1]{>{\PreserveBackslash\raggedleft}p{#1}}
\newcolumntype{L}[1]{>{\PreserveBackslash\raggedright}p{#1}}

\begin{document}
%\begin{multicols*}{2} 
\title[Towards Sustainable Creativity Support]{Towards Sustainable Creativity Support: An Exploratory Study on Prompt Based Image Generation}
%\title{One step at a time: Assessing the role of batching and partial denoising in GenAI}
%From noise to choice

%%Authors
\author{Daniel Hove Paludan}
\orcid{0009-0006-3927-9944}
\email{dpalud20@student.aau.dk}
\affiliation{%
  \institution{Aalborg University}
  \city{Aalborg}
  \country{Denmark}
}
\author{Julie Fredsgård}
\orcid{0009-0006-6652-3957}
\email{jfreds20@student.aau.dk}
\affiliation{%
  \institution{Aalborg University}
  \city{Aalborg}
  \country{Denmark}
}
\author{Kasper Patrick Bährentz}
\orcid{0009-0002-9599-5815}
\email{kbahre20@student.aau.dk}
\affiliation{%
  \institution{Aalborg University}
  \city{Aalborg}
  \country{Denmark}
}

\author{Ilhan Aslan}
\email{ilas@cs.aau.dk}
\orcid{0000-0002-4803-1290}
\affiliation{%
  \institution{Aalborg University}
  \city{Aalborg}
  \country{Denmark}
}

\author{Niels van Berkel}
\email{nielsvanberkel@cs.aau.dk}
\orcid{0000-0001-5106-7692}
\affiliation{%
  \institution{Aalborg University}
  \city{Aalborg}
  \country{Denmark}
}

\begin{teaserfigure}
\includegraphics[width=\linewidth]{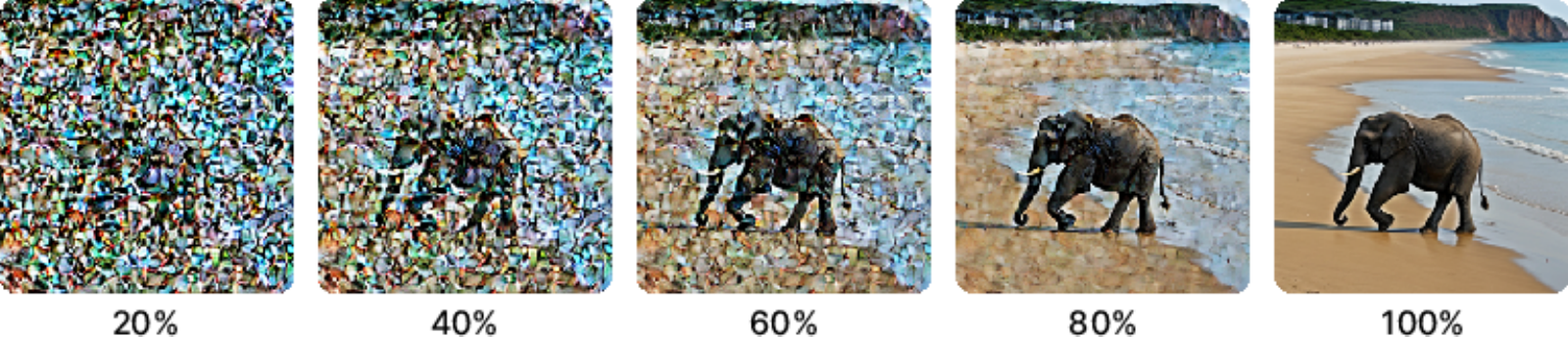}
%\vspace*{-0.4cm}
\caption{`An image of an elephant' denoised iteratively from 20\% to 100\%, showcasing how many image qualities (e.g., subject matter, composition, perspective, and colouring) are discernible before full denoising.}
%\vspace*{0.4cm}
\label{fig:denoising}
\end{teaserfigure}

\begin{abstract}
Creativity is a valuable human skill that has long been augmented through both analog and digital tools. Recent progress in generative AI, such as image generation, provides a disruptive technological solution to supporting human creativity further and helping humans generate solutions faster.
While AI image generators can help to rapidly visualize ideas based on user prompts, the use of such AI systems has also been critiqued due to their considerable energy usage. In this paper, we report on a user study (\textit{N}~=~24) to understand whether energy consumption can be reduced without impeding on the tool's perceived creativity support. 
Our results highlight that, for example, a main effect of (image generation) condition on energy consumption, and index of creativity support per prompt but not per task, which seem mainly attributed to image quantity per prompt. We provide details of our analysis on the relation between energy usage, creativity support, and prompting behavior, including attitudes towards designing with AI and its environmental impact.

\end{abstract}


\settopmatter{printfolios=true}

\maketitle

\section{Introduction} \label{Introduction}

% ________ State of the world ___________
Generative AI is finding its way into an increasing number of industries and consumer products~\cite{amato2019aimediacreativeindustries, langevin2023applicabilitydomains}, including within the creative industries of art and design~\cite{akverdi2024generative, jaaskelainen2022sustainableart}. Among the most popular products are tools for image generation, with platforms such as \textit{MidJourney}, \textit{DALL-E}, and \textit{Adobe Firefly} attracting millions of daily users. Since 2022, estimations suggest that more than 15 billion images have been generated, the vast majority ($\sim$80\%) using models, services, platforms, and applications based on Stable Diffusion~\cite{statistics2024}. Such text-to-image services allow anyone to quickly generate visual content without needing expertise in specialised software~\cite{yildirim2022text, oppenlaender2024promptingaiart}. This is further extended by recent progress in text-to-video technology, where several systems are now accessible to the public~\cite{karaarslan2024sora}. Because of this, AI tools have been described as a means of democratising art and creative production~\cite{oppenlaender2024promptingaiart} and have seen adoption in domains such as advertising and marketing~\cite{zhang2024intelligentadvertising, microsoft2024aiatwork}, urban design and architecture~\cite{yildirim2022text, paananen2023architecture, cui2024urbandesign}, and other creative fields~\cite{jian2024empoweringdesign, jaaskelainen2022sustainableart}.

% ________The big BUT_______
However, even when provided for `free', AI generation carries hidden costs due to the large amounts of energy required~\cite{luccioni2024powerWatts, varoquaux2024sustainability, chadli2024environmentalcosts, strubell2020energy}. Generating visual material emits more CO\textsubscript{2} than other AI-based tasks, with image generation consuming over 30 times as much energy as generating text~\cite{luccioni2024powerWatts}. While state-of-the-art text-to-video generators such as OpenAI's \textit{Sora} have only recently become available to the general public, the computational costs of such systems are larger still, scaling significantly with both video length and resolution~\cite{li2024carbon}. The performance of these machine learning models has improved alongside their increasing scale, but with diminishing returns. As compute demands increases faster than performance gains, this scaling approach results in an outsized environmental impact~\cite{varoquaux2024sustainability}. The large environmental footprint of generative AI (GenAI) has led researchers to call for more computationally efficient algorithms and hardware~\cite{devries2023energyfootprint}, yet few studies have explored `green AI'---that is, AI systems where efficiency is treated as a primary evaluation criterion alongside accuracy~\cite{schwartz2020green}---from an HCI perspective.

Creating a satisfactory image can be both time-consuming and costly, requiring numerous prompt iterations to generate the desired output~\cite{paananen2023architecture, oppenlaender2024promptingaiart}. This often involves a laborious cycle of trial and error, as users strive to ensure their input is interpreted in line with their intentions. Part of the challenge stems from the opacity of the underlying models, exacerbated by interfaces that lack effective feedback mechanisms~\cite{brade2023promptify, agrawala2023unpredictable}. Because the majority of generated images serve merely as stepping-stones toward an acceptable result, decreasing the amount of intermediary images---or lowering the computational costs associated with each images---would help to make image generation more sustainable. While the impact of image generators on creativity has been studied to some extent (see, for example,~\cite{paananen2023architecture,park2024howdesignersuseai}), we lack an understanding of the relation between the energy consumption and creativity support of image generators. By exploring how the design of image generators might impact the energy consumption associated with generating unused content during ideation, this paper aims to answer the research question:

%______________Therefore, we did______________ ^

\begin{quote}
    \textit{How do batching and partial denoising impact designers' use of AI image generators, and what are the implications for energy consumption and creativity support?}
\end{quote}

While batching (i.e., generating multiple images at a time) is employed by many image generators, none utilise partial denoising (i.e., the partial generation of images, as illustrated in Figure~\ref{fig:denoising}). To explore the impact of batching and partial denoising on emissions and creativity, we employed a within-subjects 2$\times$2 factorial design, where 24 participants evaluated four variants of an image generator. These differed on two variables: the level of denoising (partial vs. full) and the `prompt-to-image' ratio; i.e., the number of images generated per prompt (single vs. batch). For each combination of variables, participants were tasked with generating an image to solve one of four character design tasks. The impact on energy consumption was measured using \textit{CodeCarbon}~\cite{codecarbonmethodology}, while the impact on creativity support was measured using the Creativity Support Index (CSI)~\cite{cherry2014quantifying}.

%______________The key findings are______________
We found that partial denoising results in a significant decrease in energy consumption compared to full denoising, while single-image generation results in a significant decrease compared to batch generation. Although these variables had no significant effect on perceived creativity support at the end of tasks, interviews revealed that 20 out of 24 participants preferred batching, primarily for its ability to support exploration. The detailed analysis on the relation between energy consumption, creativity support, and prompting disclose patterns on how quality and quantity of generated images map to divergent and convergent prompting, which we illustrate and discuss. Furthermore, participants expressed a strong interest in understanding the environmental impact of their actions, suggesting the potential for incorporating eco-feedback mechanisms to inform users of their energy consumption.

\section{Related Work} \label{sec:related_work}

\subsection{GenAI in design ideation}
AI design support systems (AI-DSSs) are systems that support designers through AI technologies~\cite{lee2024aidesignprocess}. In research involving architectural, product, and UX designers, a study on the \textit{Double Diamond model} by Lee et al. revealed that AI-DSSs are able to support designers in all stages of the design process by enhancing access to information, collecting data, generating solutions, and facilitating automatic evaluations. As most existing literature covers the use of such systems in later stages of the design process (i.e., \textit{Develop} and \textit{Deliver}), the researchers call for more work concentrating on AI-DSSs in the early phases of design (i.e., \textit{Discover} and \textit{Define})~\cite{lee2024aidesignprocess}.

A study by Paananen et al. does the former, exploring the ability of three text-to-image tools (\textit{Midjourney}, \textit{DALL-E 2}, and \textit{Stable Diffusion}) to support ideation among architectural design students~\cite{paananen2023architecture}. During sessions lasting approximately one hour and 20 minutes, participants were tasked with creating the floor plan, interior, and façade of a cultural centre to explore whether image generation could be a meaningful part of the design process. Participants generated an average of 39.2 prompts, with an average sequence length of 8.4 prompts; a `sequence' being a grouping of prompts related to one idea, in which each new prompt builds upon the previous~\cite{paananen2023architecture}. This highlights how prompt engineering is an iterative process, where users continuously reject, refine, and redefine prompts until a satisfactory result is reached. While the provided tools successfully supported creativity in the ideation process, participants found it difficult to track the flow of different ideas in the vertically scrolling interfaces, suggesting that a more spatial layout would better facilitate the exploration of different design aspects~\cite{paananen2023architecture}.

In another study, Gosline et al. examined how nudging might help users identify errors in LLM-generated responses, and found that adding friction to the output review process can lead to increased accuracy---without significantly increasing task-completion time~\cite{gosline2024llmnudging}. Bach et al. similarly found that incorporating debiasing techniques into AI-powered Clinical Decision Support Systems could improve diagnostic accuracy, although clinicians expressed concerns that this might come at the expense of efficiency~\cite{bach2023allthetimeintheworld}. Such friction nudging is recommended by Weisz et al. as a general design principle for generative AI, with the aim of encouraging users to slow down and think critically about outputs at key decision-making points. This might involve offering multiple AI-generated options for users to select from or requiring users to make decisions before showing the AI’s output~\cite{weisz2024genaidesignprincipes}.

Akverdi et al. examined the impact of GenAI tools on ideation among academics and graduate students of design, technology, and society~\cite{akverdi2024generative}. Interviews exploring the disparities between AI and non-AI tools suggest that generative tools accelerate workflows, fostering creativity by providing unique insights in brainstorming sessions. While AI tools hastened decision making, concerns were raised that picking from a list of AI-generated options might restrict creativity, limiting free design and personal expression by shifting the designer's role towards one of evaluation. Effective prompt engineering also necessitates better verbal communication skills, posing a challenge for designers traditionally reliant on visual design expertise~\cite{akverdi2024generative}. Supporting this finding, a study by Park et al. conclude that current GenAI tools are ill-suited for visual thinkers and do not facilitate the non-linear nature of creative processes~\cite{park2024howdesignersuseai}. In a mapping of how professional designers use AI tools in their daily workflows, it was found that image generation systems accommodate two primary goals: sourcing inspiration and quickly translating ideas into tangible images. To better support these goals, participants call for systems that facilitate iteration (e.g., through the expansion or combination of generated images), exploration (e.g., through integrated recommender systems), and user control (e.g., through multi-model input)~\cite{park2024howdesignersuseai}.

Overall, GenAI systems demonstrate potential for expediting creative processes and supporting inspiration in the early stages of design ideation, but are limited by the need for users to communicate ideas through natural language, where the correlation between input and output is often unpredictable.

\subsection{UIs for image generation}
Prompt engineering is the practice of crafting precise and effective prompts to steer generative models toward generating specific and desired outputs~\cite{oppenlaender2024promptingaiart}. As this process entails a high degree of randomness, producing an effective prompt often involves much trial and error~\cite{brade2023promptify, agrawala2023unpredictable}. Many systems try to mitigate this by providing users additional control, through approaches such as negative prompting (specifying what the image should \textit{not} contain), prompt weighting (emphasising or de-emphasising specific words), or style selection. Some systems incorporate the design principles proposed by Weisz et al., with \textit{ChatGPT} requesting more information when a prompt is insufficiently descriptive for image generation, and \textit{Midjourney} generating four low-resolution images for users to choose from before upscaling the selected image to high definition. Other widely-used image generators, including \textit{Adobe Firefly}, and \textit{Bing}'s \textit{Image Creator} also employ this batching approach, though without allowing users to upscale individual images. \textit{Midjourney} extends this feature through `permutation prompts', where prompts are accompanied by a list of options defining possible permutations~\cite{midjourney2024permutations}. For example, the prompt "\texttt{a \{red, green, yellow\} bird in the \{jungle, desert\}}" produces six permutations, each of which generates four low-resolution images for a total of 24 images.

Recent research has explored further avenues, extending user-control through multimodal input. \textit{Promptify}~\cite{brade2023promptify}, for example, provides a drag-and-drop interface for organising and clustering generated images, providing suggestions for prompt refinement based on a selected image or cluster of images to steer the subject matter or image style in particular directions. When compared to the widely-used stable diffusion tool \textit{Automatic1111}, participants found \textit{Promptify} significantly more useful in assisting and expediting their text-to-image generation workflow~\cite{brade2023promptify}. Similarly, \textit{DesignPrompt}~\cite{peng2024designprompt} serves as a digital moodboard that sets the tone of generated images, \textit{Promptcharm}~\cite{wang2024promptcharm} provides increased insight into how prompts affect the image, while \textit{MultiDiffusion}~\cite{bar2023multidiffusion} increases user-control through spatial guiding signals such as segmentation masks and bounding boxes. Taking a similar approach, Zhang et al. shows how saliency-guided image generation gives users more control of the spatial layout of generated images~\cite{zhang2024saliencymap}. By providing a saliency map in addition to a text prompt, users are able to generate images with a specified viewer attention distribution---i.e., images that attract viewers' attention toward the desired regions---giving users increased control of image composition~\cite{zhang2023texttoimagediffusionmodels}.

Certain tools are specifically designed for inpainting, enabling users to selectively modify particular areas of an image by masking portions for regeneration. \textit{GANzilla}, for instance, facilitates the exploration of potential editing directions through scatter/gather interactions, allowing users to iteratively discover image variations and organise them into clusters~\cite{evirgen2022ganzilla}. While the system's user-driven approach led to high satisfaction among participants, the authors propose future researchers to focus on a narrower user group (e.g., product designers), and call for systems that make it easier for users to track how images change and differ across iterations.

Overall, research on user experience in image generation has focused on refining interfaces to help users better express their creative intent, emphasising clarity, timeliness, personalisation, transparency, and the minimisation of bias~\cite{amershi2019guidelinesforhumanai, weisz2024genaidesignprincipes}. This highlights the importance of aligning user input with generated outputs to ensure creativity without frustration~\cite{liu2022designguidelinesfortexttoimage}, but shows little regard for the sustainability aspects of creative processes. A growing body of work is adopting a different approach, prioritising sustainability as the primary focus in the pursuit of `Green AI'.

\subsection{Green AI}
Schwartz et al. describe Green AI as research on artificial intelligence that seeks to minimise computational resources~\cite{schwartz2020green}. Unlike `Red AI', which prioritises accuracy regardless of computational costs, Green AI focuses on optimising performance relative to efficiency and views this as an essential evaluation criteria. While there is much literature on building space- or time-efficient models, such research is often motivated by practical constraints---such as fitting a model on a small device or running it fast enough for real-time processing---rather than by considerations of environmental impact. With the goal of making AI more sustainable in all parts of its development cycle, Green AI seeks to reduce computational expense with minimal reduction in performance, and calls for researchers to report the computational costs of developing and running AI models~\cite{schwartz2020green}.

Measuring `AI waste'---i.e., the resources that are spent developing and operating AI models~\cite{reif2021aiwaste}---is a complex task influenced by various factors, including a model's parameter count, hardware configurations, and the energy efficiency of the hosting data centre~\cite{faiz2023llmcarbon}. Taking these parameters into account, \textit{CodeCarbon} measures operational emissions~\cite{codecarbonmethodology}, while \textit{LLMCarbon} projects both the embodied and operational carbon footprint of large language models, achieving high accuracy in its emission assessments~\cite{faiz2023llmcarbon}. Dodge et al. created a comparable system, concluding that choosing an appropriate region yields the greatest reduction in operational emissions. They echoed Schwartz's assertion that precise measurement is an essential first step toward minimising consumption~\cite{dodge2022measuring}.

\subsubsection{Reducing AI waste through algorithmic optimisation}
Chien et al. demonstrated that emission-aware algorithms, which route requests to low-carbon geographical locations, can reduce the carbon footprint of generative models by over 35\% without much impact on the user experience---i.e., without significant increases in response latency~\cite{chien2023reducing}. Such algorithmic optimisations can have a major impact, as evidenced by the transition from early image generation models operating in pixel space (e.g., \textit{DALL-E}) to contemporary models operating in latent space (e.g., \textit{DALL-E 2})~\cite{andrew2024stablediffusion}. This shift greatly reduces computational costs, outweighing the decoding costs required to convert images back to pixel space. As a result, resources are conserved during both training (forward diffusion) and operations (reverse diffusion)~\cite{andrew2024stablediffusion}.

While these optimisations can provide substantial energy savings, the current `bigger-is-better' paradigm of GenAI has the opposite effect. The scale of state-of-the-art models has increased massively in recent years, both in number of parameters and amount of data, exponentially increasing the necessary computing power~\cite{varoquaux2024sustainability}. Given that this development shows few signs of slowing down, finding other avenues for reducing consumption becomes increasingly relevant. While waste can be reduced at both the level of training and the level of operations, initiatives targeting operations seem most effective, as inference costs quickly outgrow training costs~\cite{chien2023reducing}. Rather than improving GenAI's underlying algorithms, some studies have accordingly turned to improving the systems, i.e., the UIs, through which users interact with these algorithms.

\subsubsection{Reducing AI waste through UI optimisation}
In the context of HCI, the principles of Green AI can extend beyond system-level improvements to influence how users engage with AI technologies. Instead of solely focusing on how researchers can optimise models, efforts are being made to raise awareness and encourage more sustainable AI practices among users. For example, `The Green Notebook' explored how artists could reflect on the sustainability of AI tools in their creative practices. By offering participants a platform to discuss and receive feedback on the environmental impact of their ideas, Deng et al. found that many artists are unaware of the carbon footprint associated with AI tools and struggle to assess the impact of their work. However, participants responded positively to tools designed to enhance their understanding and minimise their energy consumption, which in turn increased interest in the sustainability of their creative processes~\cite{deng2023green}.

This aligns with the research of Dodge et al., who argue that providing users with information about software carbon intensity is a crucial first step toward reducing emissions~\cite{dodge2022measuring}. Multiple generative systems have been designed to provide this sort of `eco-feedback'~\cite{froehlich2010ecofeedback}, with interfaces supporting and promoting more sustainable interaction. Ren et al. found that integrating sustainability reflection features into GenAI systems was an effective way of communicating energy consumption~\cite{ren2023guiltyinagoodway}. They discovered that presenting consumption information in relatable terms---such as equating energy use to watching \textit{X} hours of \textit{YouTube}---was an effective method for delivering feedback on energy consumption. For the majority of participants, increased awareness caused a desire to go back and redo their process, using lower consuming settings during initial experimentation or drafts~\cite{ren2023guiltyinagoodway}. Such eco-feedback can be described as a type of `disclosure nudge', i.e., a nudge that adds relevant information to the choice the user is about to make. Other forms of nudging, including feedback nudges, default nudges, and social nudges, have likewise demonstrated potential in promoting sustainable consumption behaviour~\cite{lehner2016sustainablenudging}.

\begin{figure}[ht!]
  \includegraphics[width=.6\linewidth]{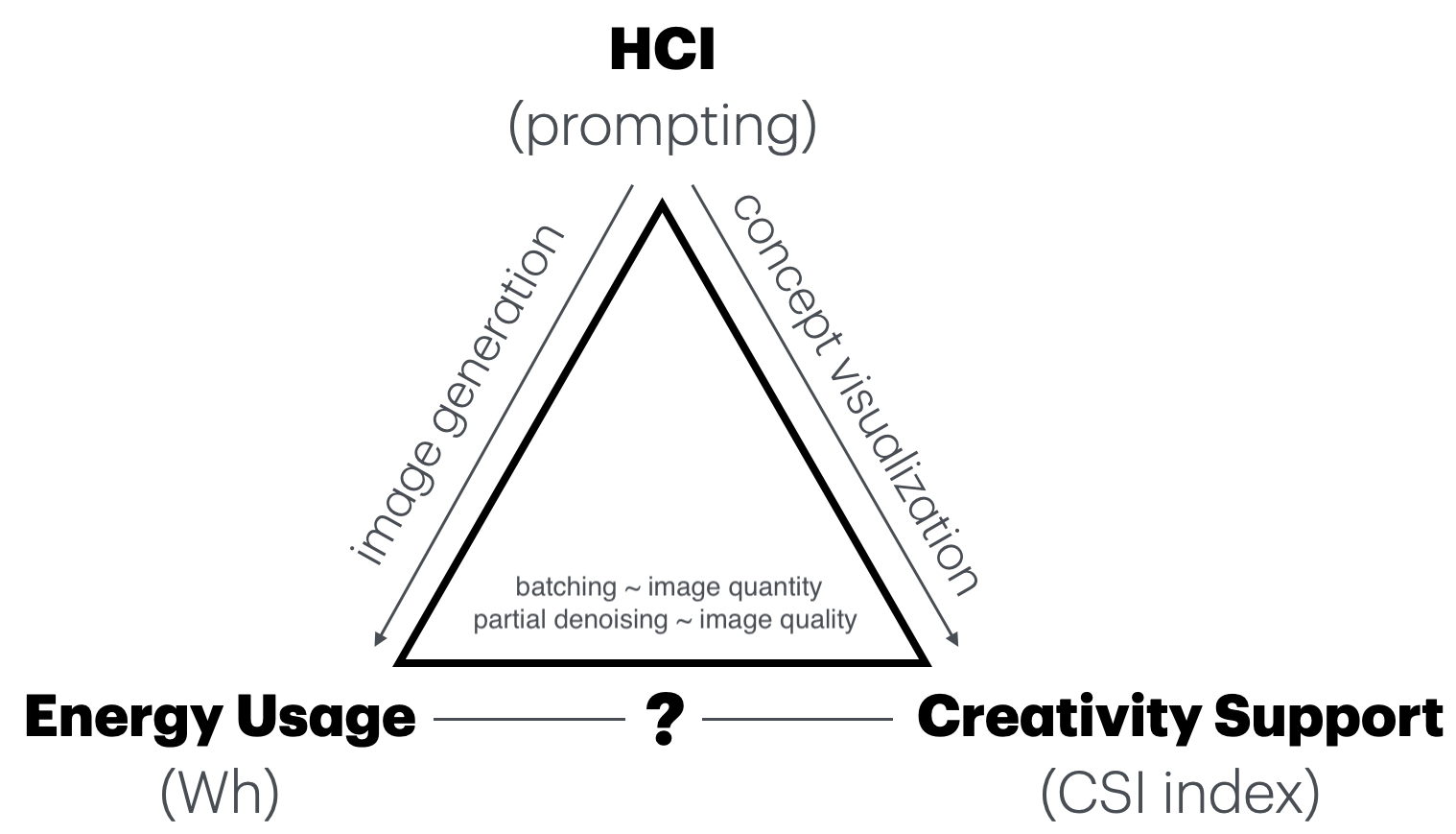}\\
  \caption{Overview of our research concept. We aim to better understand the relation and balance between \emph{energy usage} and \emph{creativity support} through the study of effects of quality and quantity of images generated to users and their \emph{prompting} (interaction).}
  \label{fig:research_concept}
\end{figure}

In general, the pursuit of green AI presents challenges both on a technical and a human-centred level. Although substantial algorithmic improvements have been made to reduce the waste of GenAI systems, there is a lack of research on how the user-facing side of these systems can be designed to support sustainable usage. To bridge this gap, this paper explores how two components of image generators---denoise level and prompt-to-image ratio---impact their energy consumption and ability to support creativity (see Figure~\ref{fig:research_concept}).

\section{Method} \label{sec:method}
\begin{figure}[ht!]
\includegraphics[width=.6\linewidth]{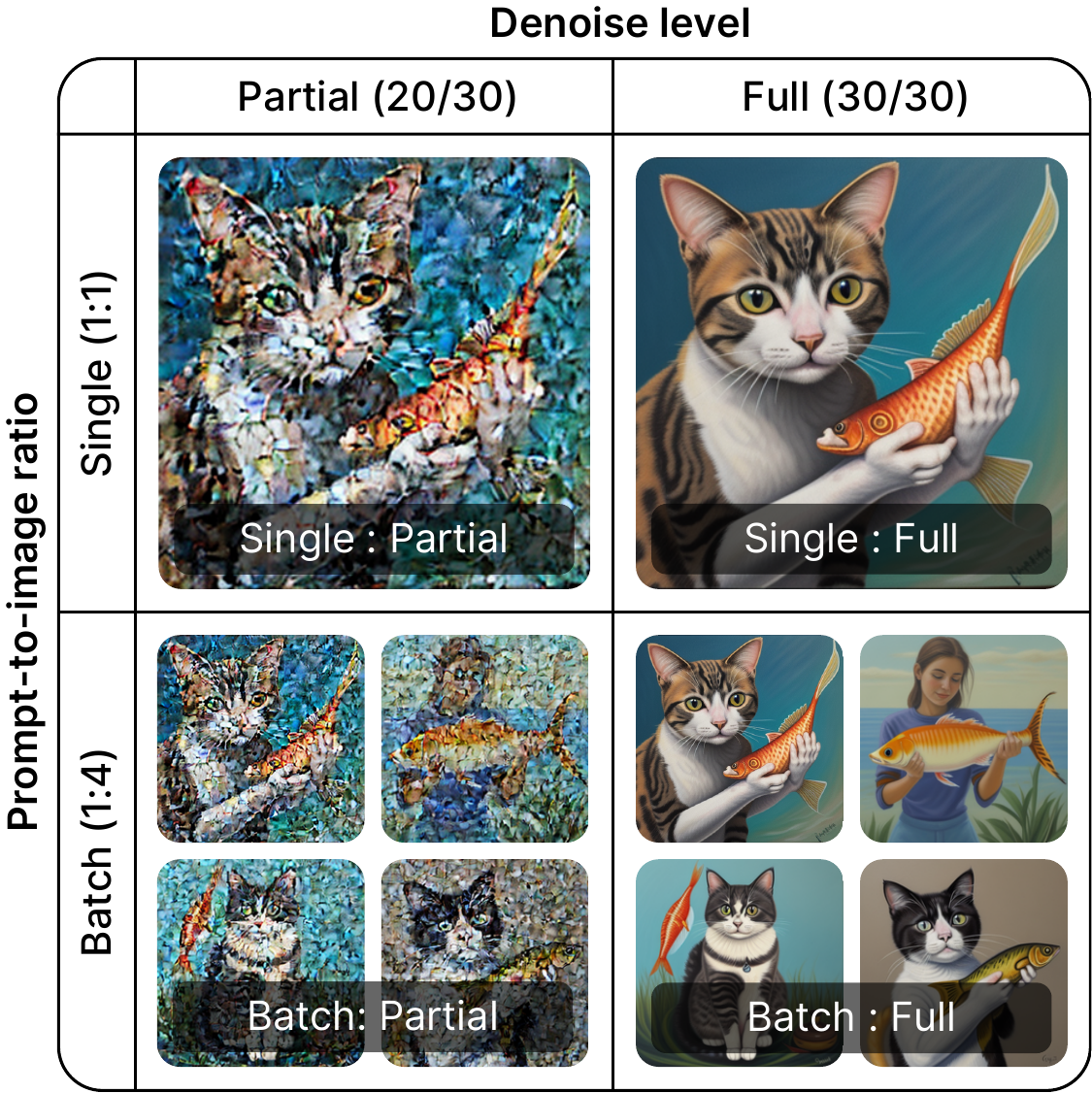}
\caption{Example of the four conditions in the 2$\times$2 factorial design for the prompt: `A cat holding a fish'.}
\label{fig:conditions}
\end{figure}

To explore the effects of image quantity (i.e., prompt-to-image ratio) and  image quality (i.e., denoise level) on energy consumption and user perception, we used a within-subjects 2$\times$2 factorial design, testing two independent variables: \textit{the level of image quality} (partial vs. full) and \textit{image quantity} (single vs. batch). 
As shown in Figure~\ref{fig:conditions}, these categorical dimensions combine to four conditions:

\begin{itemize}
    \item \textbf{Single} image : \textbf{Partial} denoising
    \item \textbf{Single} image : \textbf{Full} denoising
    \item \textbf{Batch} of images : \textbf{Partial} denoising
    \item \textbf{Batch} of images : \textbf{Full} denoising
\end{itemize}

%To evaluate these, participants interacted with four variants of an image generator. %The system we designed for the evaluation is described in detail in Section~\ref{sec:method}.

%\subsection{Design a Character Task}
To evaluate these conditions, we created a prototype image generator and asked participants to perform design tasks with four variants of this generator. For each condition, we assigned participants one of four design tasks; to design a character for a: (i) children's story, (ii) horror story, (iii) superhero story, and (iv) fantasy story. These tasks were selected for their tangible, open-ended nature, and low entry barriers---all participants had a basic understanding of what a character in these genres might look like, enabling them to generate ideas regardless of their academic background.
To complete the tasks satisfactorily, participants were encouraged to complete each task within a five minutes frame and design their characters in detail by addressing the following questions: (i) What do they look like? (build, facial features, etc.), (ii) What are they wearing? (clothing, accessories, etc.), (iii) Where are they? (location, time period, etc.), and (iv) What are they doing? (behaviours, emotions, etc.).

\subsection{Hypotheses}
We evaluated four hypotheses; two pertaining to energy usage and two pertaining to creativity support:

\begin{itemize}
    \item [\textbf{H\textsubscript{1.1}}] Energy consumption is higher when completing a task in the full than partial denoise conditions.
    \item [\textbf{H\textsubscript{1.2}}] Energy consumption is higher when completing a task in the batch than single image conditions.
    \item [\textbf{H\textsubscript{2.1}}] CSI scores are higher when completing a task in the full than partial denoise conditions. 
    \item [\textbf{H\textsubscript{2.2}}] CSI scores are higher when completing a task in the batch than single image conditions.
\end{itemize}

H\textsubscript{1.1} and H\textsubscript{1.2} were formulated based on our knowledge that fully denoising an image consumes more energy than partially denoising it, and that generating a batch of images consumes more energy than generating a single image.

As both batch generation and full denoising lead to significantly higher emissions per prompt, we hypothesise the same to be true per task. This inference is based on two assumptions: first, that the total amount of generated images is significantly higher when generating in batch; and second, that the total amount of generated images is comparable between the partially denoised and fully denoised conditions. In both cases, the net effect is an increase in energy consumption, either due to an increase in the total number of images (batching) or due to an increase in the cost of each image (full denoising).

H\textsubscript{2.1} and H\textsubscript{2.2} were formulated based on the assumption that fully denoised images provide clearer visual guidance for creative tasks, and that batch conditions better facilitate inspiration gathering through variety and comparison, thus providing better creativity support.

\subsection{Participants}
24 participants (17 male, 7 female) were recruited among design students from (blinded for review) University and (blinded for review) College. Half of the participants were interaction design students, the rest studied graphical design, industrial design, data science and machine learning, interactive digital media, and software engineering. All participants had studied in their respective programmes for at least two years. A fourth of participants were between 18 and 24 years of age; the rest between 25 and 34.

\begin{figure}[ht!]
  \centering
  %\subfloat{\includegraphics[width=8.5cm]{graphics/study_setup.jpg}}\\
  %\vspace{-0.6em}
  \subfloat{\includegraphics[width=8.5cm]{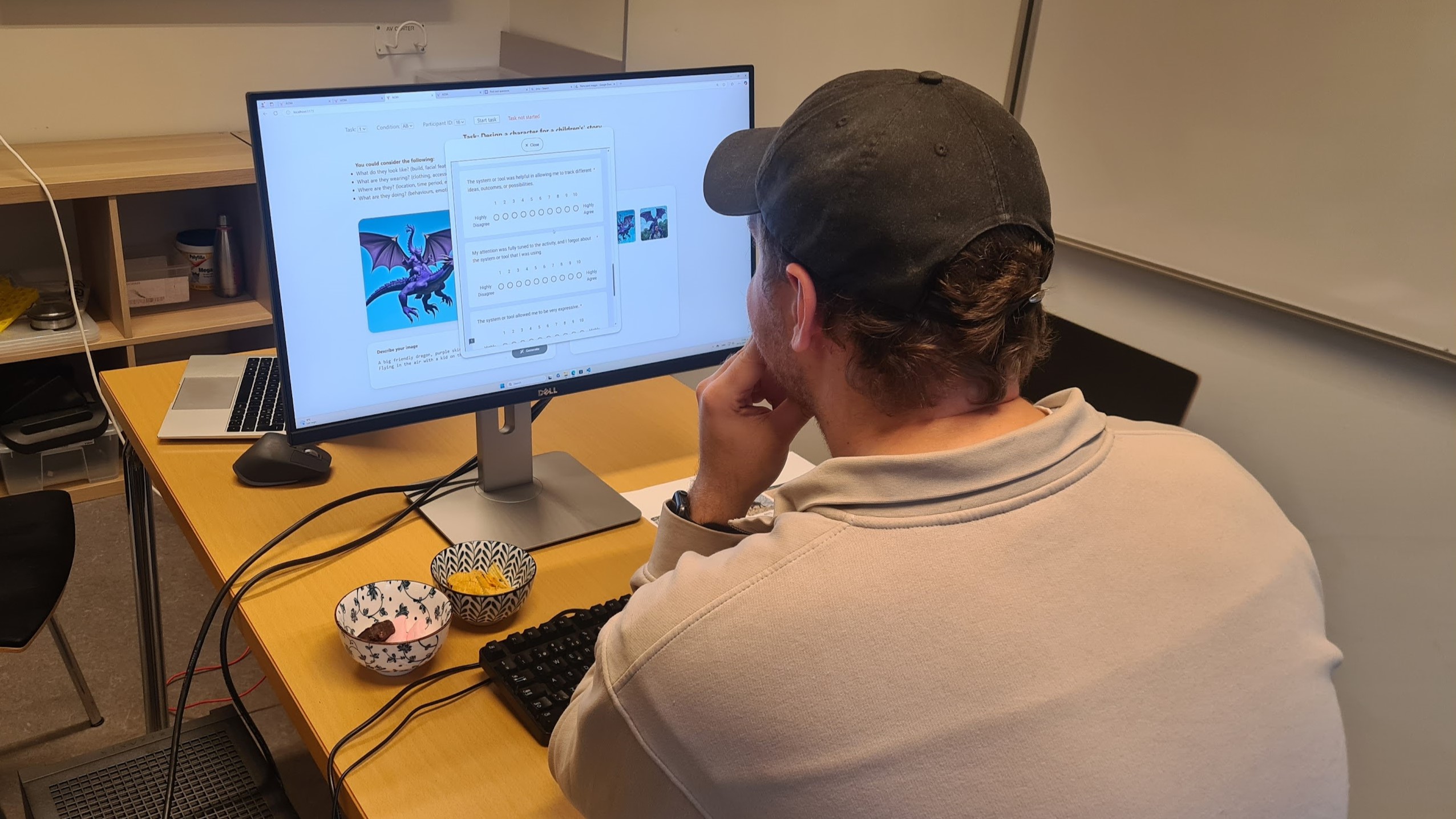}}\\
  \caption{Setup of the user study showing a participant answering a CSI questionnaire after trying his third condition. }
  \label{fig:study_setup}
\end{figure}

\subsection{Apparatus and Procedure} \label{sec:system}
To evaluate the impact of denoise level and prompt-to-image ratio on energy consumption and creativity support, we developed a web-based prototype. The prototype implements the four experimental conditions, enabling participants to interact with each condition while completing the specified tasks. It was developed using \textit{Python} for the backend and \textit{React} for the frontend. The system employed the Stable Diffusion 1.5 based \textit{Juggernaut Reborn} model~\cite{juggernaut}, running locally on the computer. Due to limitations to the model, prompts were capped at 160 characters. The UI interface was kept simple with the main interaction happening on the left side with an image view (depending on condition) and the prompting window. On the right side participant could see their history of generated images.

For this evaluation, we set a fully denoised image to 30 steps, as recommended in the documentation of the model~\cite{juggernaut}. The partially denoised images had completed 20 out of the 30 steps, as we deemed that these were sufficiently clear to identify the contents of the image. 

Evaluations were conducted over four days at two locations within (blinded) University. Participants were seated in front of a monitor equipped with a keyboard and mouse, as shown in Figure~\ref{fig:study_setup}. 

To address potential carryover effects, such as learning and fatigue, between tasks all possible orders of conditions ($4! = 24$) were tested across the 24 participants. Tasks were distributed in the same way, testing every possible order. Figure~\ref{fig:denoising} was used to introduce participants to the concept of denoising at the beginning of the task. To not bias participants behaviour we chose not to inform the participants that the study was about sustainability.

Following each condition, participants completed a questionnaire assessing how well the respective system performed on five dimensions of the Creativity Support Index: \textit{Exploration}, \textit{Expressiveness}, \textit{Immersion}, \textit{Enjoyment}, and \textit{Results Worth Effort}\footnote{The \textit{Collaboration} dimension was excluded, as it was deemed irrelevant to this evaluation focusing on individuals}. The CSI is a survey designed to evaluate the effectiveness of creativity support tools (CSTs) in assisting users engaged in creative tasks. We consider image generators CSTs, as their ability to generate, refine, and communicate ideas fits the definition of tools that ``\textit{can be used by people in the open-ended creation of new artifacts}''~\cite{cherry2014quantifying}. The CSI has previously been used for similar evaluations, particularly by Paananen et al., who applied it to examine the role of text-to-image generation in architectural design ideation~\cite{paananen2023architecture}.

The amount of energy consumed by each participant and each condition were measured using the \textit{Python} package \textit{CodeCarbon}, which tracks the power consumption of the device and calculates the carbon intensity of the electricity consumed based on the mix of energy sources in the local grid~\cite{codecarbonmethodology}. Additionally, each participant's number of prompts and generated images were recorded for every condition.

Having completed all conditions and their corresponding questionnaires, participants were presented with a final questionnaire. In this questionnaire, participants filled out the CSI subfactor comparison along with questions that explore users' preferences and attitudes toward using AI tools in terms of creativity, efficiency, and environmental impact. Finally, follow-up interviews explored participants' experiences with each of the four systems. The short (>10 minutes) semi-structured interviews focused on the perceived pros and cons of each condition across tasks, with particular attention to how partially denoised images affected participants' sense of control, creative approach, and understanding of the underlying processes.

\subsection{Analysis}
 CSI scores were calculated using the method described by Cherry and Latulipe~\cite{cherry2014quantifying}.
The statistical analyses were performed in \textit{R} using mixed-effects models followed by Tukey post-hoc tests for pairwise comparisons. Additionally, linear regression was used to examine the relationship between CSI scores and energy consumption. 

To analyse the interviews, we employed thematic analysis. Initially, all 24 interviews were transcribed. Subsequently, we divided the interviews among team members, with each member independently reviewing eight interviews and highlighting notable quotes. After this initial phase, we rotated, ensuring that each interview underwent review by at least two team members. A total of 226 quotes were selected for the subsequent coding phase. During this stage, we collectively sorted related quotes into codes, which were then clustered into broader categories. This iterative process continued, culminating in the final categories and themes presented in Table~\ref{tab:themes} and detailed in the following section.

\section{Results}
 We first report on quantitative data analysis  to test our hypotheses, followed by our exploratory and qualitative analysis to contrast the hypothesis-driven results and to uncover the factors driving the observed outcomes. The quantitative analysis is based on a range of measures, including \textit{CodeCarbon} data, the results of the CSI and post-test questionnaires, and statistics on participants' usage of the system.

\begin{figure}[ht!]
  \centering
  %\subfloat{\includegraphics[width=8.5cm]{graphics/study_setup.jpg}}\\
  %\vspace{-0.6em}
  \subfloat{\includegraphics[width=.99\linewidth]{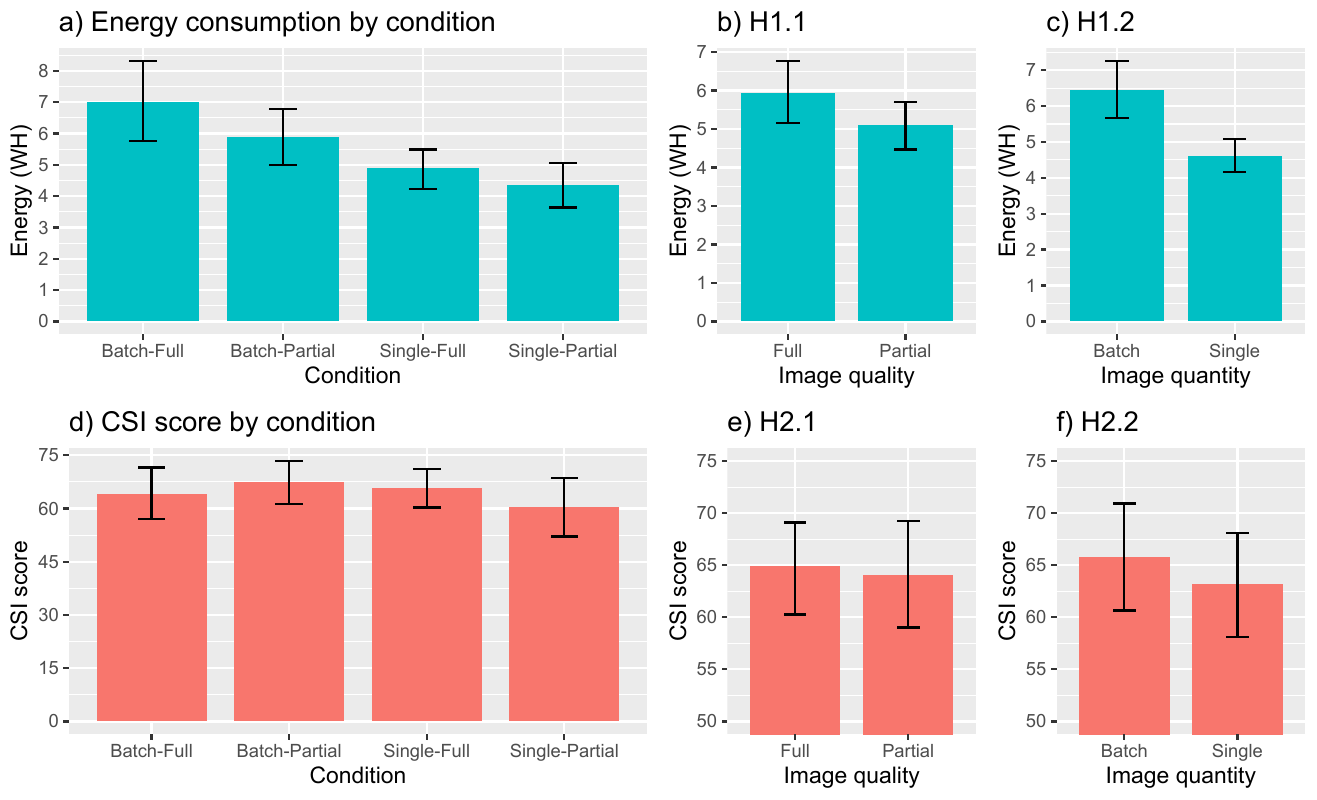}}\\
  \caption{Visualization of main ratings and data per task (i.e. CSI ratings, energy consumed).a) and d) show CSI score and energy consumption for the four conditions b) relates to H\textsubscript{1.1}, b) to H\textsubscript{1.2}, e) to H\textsubscript{2.1}, and d) to H\textsubscript{2.2}. f). Error bars denote 95\% confidence intervals.}
  \label{fig:QQ}
\end{figure}

\subsection{Quantitative analysis}
 \subsubsection{Hypotheses testing}
 Considering H\textsubscript{1.1} and H\textsubscript{1.2} (see Figure~\ref{fig:QQ}b and c), we found  that image quality defined by the two denoise levels (i.e., partial vs full) had a main effect on energy consumption with the effect being significant ($\chi^2 \left(1\right) = 4.04, \, p = .044$). Tukey post-hoc analysis revealed that user interactions consumed significantly more energy in tasks where users had to work with fully denoised images compared to the tasks where users had to interact with partial denoised images ($z = 2.04, p = .042$), with an average difference of approximately 0.84 Wh. These results support H\textsubscript{1.1}. The analysis also showed that the prompt-to-image ratio significantly affected energy consumption ($ \chi^2 \left(1\right) = 13.96, p < .001 $). Tukey contrasts indicated that users consumed  significantly more energy when generating \emph{batch} of images compared to a \emph{single} image   ($z = 4.35, p < .001$), with an average difference of around 1.8 Wh. These findings support H\textsubscript{1.2}. 
Considering H\textsubscript{2.1} and H\textsubscript{2.2} (see Figure~\ref{fig:QQ}e and f), we found no main effect of denoising level (i.e., partial vs full) or prompt-to-image ratio (i.e., single vs badge) on the CSI score (per task) with ($p=.7626$) and ($p=.3558$). Considering Figure~\ref{fig:QQ}d statistical tests revealed also no significant main effect of condition (i.e., \emph{Batch-Full}, \emph{Batch-Partial}, \emph{Single-Full}, and \emph{Single-Partial}) on CSI score ($p=.3558$).

Figure~\ref{fig:QQ} a and b show mean ratings for each condition and task. We can assume that the energy consumption task relates to the energy consumption per prompt for each condition.
The variation between users CSI scores in each condition shows low variations.  
These results suggest that users interaction behavior must have been somewhat similar considering number of prompts written and task completion times. Furthermore, users mean ratings show little variations considering creativity support over tasks and conditions.  
In the following we will explore these outcomes and user interaction behavior in more detail.

\begin{figure}[ht!]
  \centering
  \subfloat{\includegraphics[width=.99\linewidth]{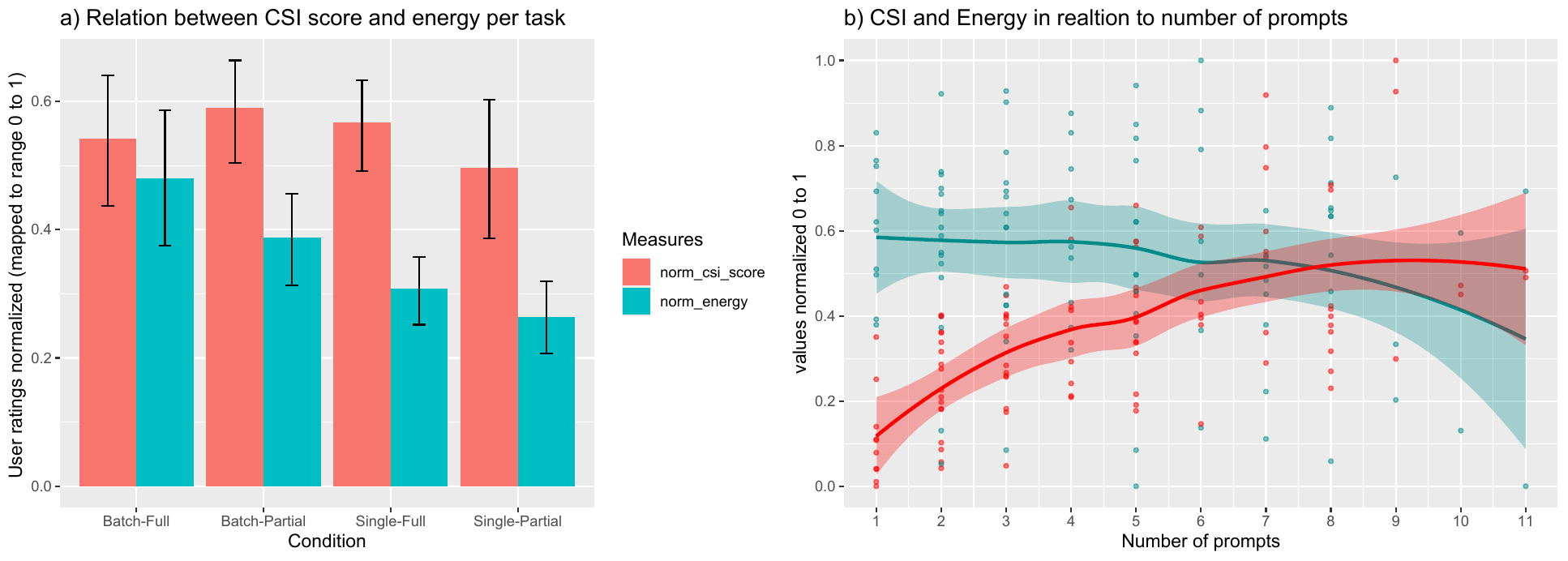}}\\
  \caption{Overview of user CSI ratings and energy consumption showing a) for each condition the relation between CSI rating and energy per task, and b) how the ratings and energy consumption develop in relation to the number of prompts per task. Values have been normalized (i.e. mapped to the range 0 to 1). Error bars denote 95\% confidence intervals.}
  \label{fig:CSI_Energy_norm}
\end{figure}
\subsubsection{Exploratory analysis}
Both the number of prompts written during a task and the duration of time used to complete the task correlate to an increase in energy consumption ($p<.001$ each). Analysis further indicates a negative correlation between CSI score and energy consumption ($p=.0048$), suggesting that systems which better support creativity result in lower consumptions overall. However, while statistically significant, linear regression analysis indicates only a weak relationship between these variables ($R^2 = .082$). The correlation may thus be indirect, potentially driven by the negative correlation between task duration and CSI score ($p<.001$). Evidence for this interpretation is provided by the negative correlation between duration and \textit{Results worth effort} ($p<.001$), underscoring the importance of task completion time for creativity support. 
Using the same linear mixed-effects model method as mentioned above, we found a significant difference in CSI per prompt for the four conditions ($p=.0401$). However, post-hoc tests revealed only a marginally significant difference ($p=.0587$) between the \textit{Batch of images : Partial denoising} condition and the \textit{Single image : Full denoising} condition, with an estimated mean difference of 7.92 ($SE = 3.16$, $z = 2.51$, $p = .059$). This suggests that the \textit{Batch of images : Partial denoising} condition needs fewer prompts to reach the same CSI score as the \textit{Single image : Full denoising} condition.

\begin{figure}[ht!]
  \centering
  \subfloat{\includegraphics[width=.99\linewidth]{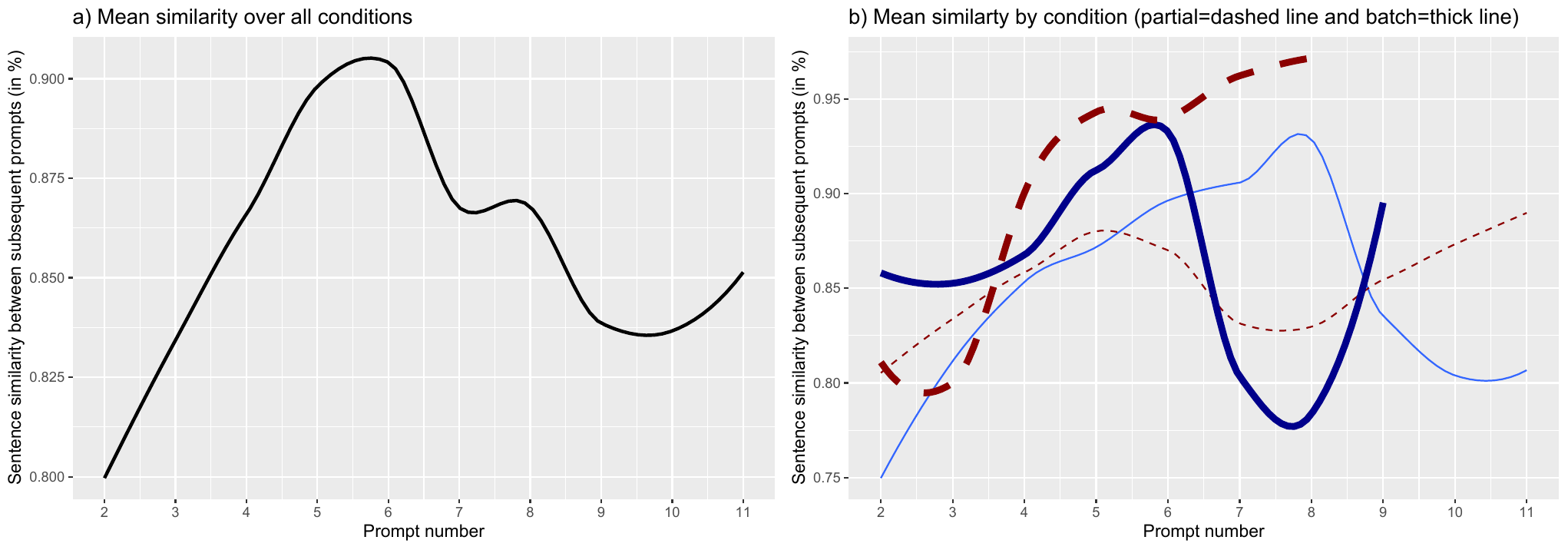}}\\
  \caption{Overview of how prompts change over time measured by sentence similarity of subsequent prompts, indicating episodes of convergence and divergence during the ideation.}
  \label{fig:prompt_similarity}
\end{figure}

To better understand the prompting behaviour we measured distance (or similarity) between subsequent prompts. To this end we utilized a pretrained `SentenceTransfromer' model~\cite{reimers-2019-sentence-bert} and sentence embeddings. Doing this, allowed us to visualize when participants converged in a prompt and idea and when they diverged. Figure~\ref{fig:prompt_similarity} visualizes this information in a) overall conditions, showing that the fist five to six changes in the prompt are performed to iteratively fine adjusting the prompt and either achieving success or starting to diverge from the prompt and start over. In b) we illustrated the same information separately for the different conditions; we can see that for the batch-partial condition (thick and dashed line) convergence happens with less prompts and overall less prompts are used.

\begin{figure}[!ht]
\includegraphics[width=.5\linewidth, trim=0 0 0.5cm 0, clip]{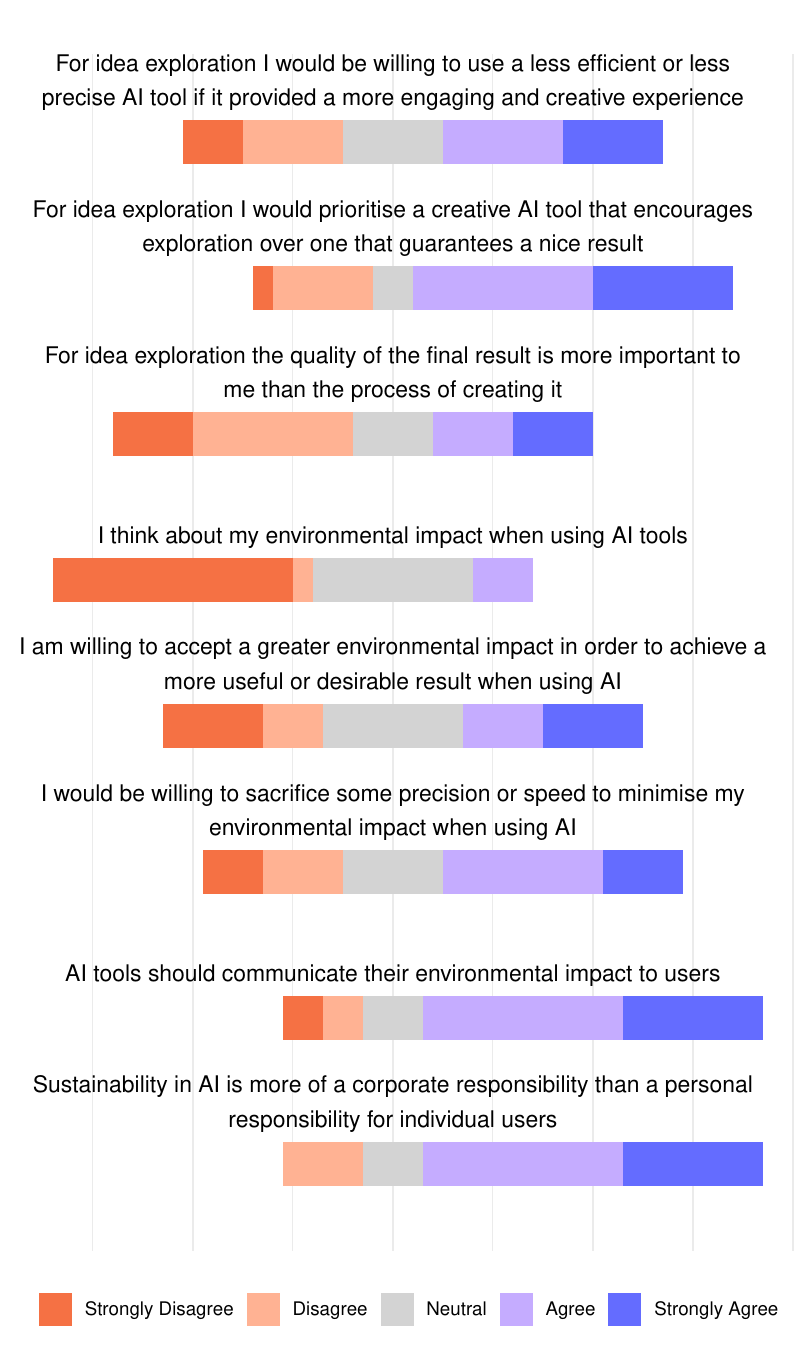}
\caption{Likert scale responses for the post-test questionnaire on AI in design processes and AI sustainability.}
\label{fig:questionnaire}
\end{figure}

\subsubsection{Creativity and sustainability questionnaire}
Figure~\ref{fig:questionnaire} shows the results from the post-test questionnaire.

On creativity, the majority (16) of participants expressed a preference for AI that supports creativity over AI that guarantees a nice result, while half report that the process of creating an image is more important than the image itself. Supporting these numbers, many (11) state a willingness to use a less efficient or precise tool in trade for a more engaging and creative experience. 

Regarding sustainability, the data reveals that a majority (13) of participants do not consider environmental concerns when using AI tools; most (17) feeling that this is primarily the responsibility of corporations. However, most (17) also believe that AI tools should clearly communicate their environmental impact to users, and half expressed openness to sacrificing precision or speed to reduce this impact. This highlights a desire for transparency in regard to energy consumption, even if users are not actively considering environmental factors themselves.

\subsection{Qualitiative analysis} \label{QualitativeAnalysis}
The qualitative analysis is based on the conducted post-test interviews.
Through thematic analysis, we identified four key themes spanning nine categories, as summarised in Table~\ref{tab:themes}. Two themes reflected participants' attitudes towards the experimental variables (denoise level and prompt-to-image ratio), while the other two highlighted participants' suggestions for increasing the sustainability of AI tools and designing image generation systems that better support ideation\footnote{All quotations have been translated into English}.

\begin{table*}[!ht]
\begin{tabular}{llll}
\hline
{\textbf{Quotes (\textit{N})}}             & \textbf{Categories}                      & \textbf{Themes}                 \\ \hline
\addlinespace
74                               & Interpreting noisy images                & Noise and creativity support    \\
                                    & Speed and quality                        &                                 \\
                             & Communicating noise to the user          &                                 \\
\addlinespace
34                            & Options support creativity               & Batching and creativity support \\
                                   & Single images are easier to work with    &                                 \\
                                      &                                          &                                 \\
\addlinespace
32                                     & AI as a tool                             & Using AI for ideation           \\
                               & Features supporting creativity           &                                 \\
                                       &                                          &                                 \\
\addlinespace
86                                      & Impact on behaviour                      & Sustainability and AI           \\
                             & Integrating sustainability into the tool &                                 \\
                                             &                                          &                                 \\
\end{tabular}%
\caption{Overview of thematic analysis findings from the study interviews.}
\label{tab:themes}
\end{table*}

\subsubsection{Noise and creativity support}
During the interviews, nine participants stated a preference for conditions with partially denoised images, while the remaining 15 participants preferred conditions with fully denoised images. 
Some participants argued that fully denoised images better support creativity: ``\textit{I was just interested in seeing the clear [images] to be able to spark more ideas.}'' (P20). Others preferred partially denoised images for their creative process: ``\textit{I actually preferred it when it was blurry rather than when I saw the final finished product [...]. It did not look as good when it was finished, but when it was blurry, it helped to just create new ideas and things like that, so it was better.}'' (P11).

Participants also disagreed on whether images with noise were adequate for developing ideas. Many found it frustrating not being able to clearly see the initial image and having to go through extra steps to fully denoise it: ``\textit{It was also annoying because I could not see what I had in the beginning, so I had to press continue.}'' (P4). In contrast, some participants found partially denoised images sufficient for assessing whether an image was heading in the desired direction: ``\textit{If I just need an idea about whether they should wear a dress, or what colour it should be, or its shape, then I might as well see it here, I do not need more steps.}'' (P1). Some participants added that it felt faster, and they therefore found it preferable to work with the noisy images: ``\textit{If I had to say which one I liked the best, it would probably be the first one where it created a slightly grainy image initially. That was probably because it was produced a bit faster.}'' (P21). 

We deliberately chose not to explain the purpose behind partial denoising to participants, aiming to gather their unbiased opinions. Interestingly, multiple participants expressed a more favourable view of partial denoising after learning that its purpose was to reduce energy consumption. For instance, P23 initially found partial denoising limiting to her creativity, but upon discovering its energy-saving intent, she remarked, ``\textit{If it was communicated to me why the images were blurry, I think that would change my opinion on it.}'' (P23). This underscores the difficulty in fully understanding participants’ true attitudes, as their initial reactions may not reflect how they might respond with more contextual knowledge.

Despite the partial denoise level being consistent at 20/30 in both of the partial denoise conditions, some participants found it easier to discern details when noise was applied to a batch of images, rather than to a single image: ``\textit{Now I am thinking more in terms of noise, the amount of it, which feels nicer here than the last one. I am not sure if there is a difference, but here (}Single image : Partial denoising\textit{), it feels a bit annoying, whereas in the first one (}Batch of images : Partial denoising\textit{), it feels like it made sense.}'' (P2). A possible explanation for this difference could be that the four images in the batch condition were presented at a smaller size compared to the single large image. This might have made it easier for participants to focus on the overall composition and less on minor details, providing a clearer sense of the image as a whole.

Participants were introduced to the denoise process using Figure~\ref{fig:denoising} with a brief explanation of the stepwise image generation process. While most participants navigated the program correctly, some were confused about the progress bar's relationship with the \textit{Continue} button. For instance, P10 noted that the lack of clear indicators, such as percentages, made it difficult to understand how much progress was made after each click.

\subsubsection{Batching and creativity support}
During the interviews, 20 participants expressed a preference for conditions where they were presented with four images, while only four participants preferred conditions with a single image. The reasons for favouring four images over one were quite consistent. Many participants emphasised that having four images provided more options for comparison, which was beneficial in their creative process: ``\textit{I think I liked those that gave me four at a time the best [...] The ones where I only got one, I ended up sitting and clicking a lot more times than I did with four. I really liked getting a lot at once because I wanted something to compare with the others.}'' (P13). Two participants even suggested being presented with more than four images.

Participants who preferred the single image conditions argued that having four images could be somewhat confusing and that dealing with just one image was easier: ``\textit{I actually think I preferred the ones where there was just one. [...] I found it difficult to move on from one out of four. Whereas when there was only one image, I felt like it made sense, okay, what is the next thing I want to add or remove.}'' (P2). 

In addition to preferences around conditions, participants were also focused on task-specific outcomes. When asked which version or elements of the four different programs they preferred, seven participants first referred to specific tasks---prior to discussing their opinions on noise and batching---and stated that they preferred certain versions because the results better matched their expectations or allowed them to achieve a satisfying result within the limited five-minute time frame. For example, Participant 14 stated: `\textit{I actually preferred this one where I had to create a character for a horror story; that is where I best could get what I asked for.}'' (P14). This suggests that participants might rate the tool more favourably if they achieve their desired results more quickly, potentially linking this to the correlation between CSI and task completion time/energy usage.

\subsubsection{Using AI for ideation}
Almost all participants had prior experience with AI image generation, though only a few had used it specifically for ideation. Drawing on their broader experiences, participants suggested various enhancements to the tool's ideation capabilities. Many of the suggestions reflected features they had encountered and appreciated in other image generation programs, which we deliberately excluded from our program to maintain focus on the experimental variables. Many expressed interest in being able to iterate on images by submitting follow-up prompts or using inpainting to modify specific areas. Additionally, some participants desired more adjustable parameters at the outset, such as options for style or tone.

Many participants hit the 160 character prompt cap, explaining how this limited their creativity: ``\textit{What held me back a bit, also in terms of idea generation, was perhaps the length of the prompt [...] I had to limit myself. I think I spent a lot of time figuring out how to phrase my sentence---what to add or remove---so it would fit within the constraints.}'' (P21). While one participant mentioned how this constraint could facilitate creative thinking, others agreed that longer prompts would benefit their creative process.

P3 proposed that the system should assist users with prompt engineering by offering examples to achieve specific outcomes, such as particular poses. P23 recommended incorporating multi-modal prompts inspired by mood boards, enabling users to combine diverse ideas---potentially resembling the approach used in \textit{DesignPrompt}~\cite{peng2024designprompt}. P11 suggested visualising the ideation process as a tree, with the branches representing various directions an idea could take.

\subsubsection{Sustainability and AI}
The post-questionnaire revealed that most participants did not consider the environmental impact of AI tools, believing this is primarily the responsibility of the corporations behind the systems (see Figure~\ref{fig:questionnaire}). During the interviews, participants provided further insight into their responses, citing various reasons for their views.

Some were unaware of AI's environmental impact, with one participant stating: ``\textit{I did not know that it impacted the environment at all, so no, I have not thought about it.}'' (P8). Others acknowledged their awareness but admitted to either forgetting, not caring, or not considering it their responsibility. For example, one participant said: ``\textit{I do know that when I prompt, it is somewhat of a burden on the environment. And I have to be honest and admit that it is not something I think about a lot when I am using it myself.}'' (P23), while another said: ``\textit{I do not really care about the environmental aspect because I do not think it should be my responsibility for this tool to be sustainable; it is [the responsibility of] whoever provides it.}'' (P3). 

However, one participant offered a contrasting viewpoint, suggesting that AI could sometimes save resources by increasing efficiency: ``\textit{When you, for example, use AI to optimise things, you often end up actually improving things, which can reduce environmental impact [...] So, there is also a positive side to it sometimes, which one could consider as well.}'' (P11).

When asked whether they were interested in knowing their environmental impact when using AI tools, most participants agreed that it would be nice to be informed about: ``\textit{I definitely think it has some potential. I actually believe it could do some good. It could potentially also help create a kind of... at least just awareness about how much energy AI actually uses.}'' (P21). A few participants expressed disinterest, citing potential stress and worsened user experience: ``\textit{I think it is a good idea you have, because there are probably some who would use it [...] Personally, I would just find it a bit stressful to know every time I generate something, that it says I have polluted so much.}'' (P9). This sentiment aligns with multiple statements suggesting that if such information were to be implemented, it should be informative rather than accusatory: ``\textit{It is very important that it does not become like `you are doing this and this and this wrong', but that somehow you manage to present it in a way where you are just informed, and then you can decide for yourself.}'' (P12). Many participants also found environmental measurements like CO\textsubscript{2} or energy use abstract, proposing that relating these to more familiar concepts could improve understanding: ``\textit{Maybe [do] not say `so many kilowatts' but like ´with this prompt you could actually have charged your phone 10\%', or something, so you get more context on how much you are actually using.}'' (P15).

Despite interest in being informed about their impact and providing suggestions on presentation, participants were generally uncertain whether this information would actually influence their behaviour. For example, participant 11 stated: ``\textit{I think it would make a difference if people were shown how much computing power it takes to run these things.}'' (P11). However, when asked if this would make a difference for him personally, he responded: ``\textit{For me? No, probably not.}'' (P11). \\

\section{Discussion}
This study explores the relation between prompting behavior, energy usage, and creativity support as consequences of the interaction. While creativity support is desirable, too much energy consumption is a ``more than human'' design issue \cite{Giaccardi04072025}. Our research has shed some light on this dilemma. The data supports the assumption that participants use more energy when completing a design task with batching and fully denoising. This is not surprising, even though one could have made the argument that showing more images at a time or higher quality images to users could potentially result in energy-efficient behavior, such as prompting less often. In this context, the prompting behavior we report in Figure \ref{fig:prompt_similarity} is potentially a valuable contribution to help optimize prompting behavior, considering a mix of batching and denoising strategies during the same task. Also, the significant results regarding CSI per prompt have potential to open new research directions. While no significant differences were measured regarding CSI scores per task, the conducted interviews revealed a preference among participants for both batching and full denoising. Employing batching aligns with Weisz et al.'s design principle of providing options,~\cite{weisz2024genaidesignprincipes}, and the preference for this could be attributed to the added flexibility and inspiration provided by multiple options, with two participants expressing a desire for more than four images per batch. However, as discussed by Iyengar and Lepper~\cite{iyengar2000choice}, while having sufficient choices is crucial for meaningful decision-making, an abundance of options risks overwhelming users and reducing satisfaction. The preference for full denoising might stem from a more frictionless experience that allowed participants to focus on their creative goals without interruptions. This aligns with broader user experience principles, and the notion that minimising friction is key to creating satisfying interactions.
Partial denoising, while less popular, was still appreciated by several participants. Some found the slightly blurry, partially denoised images helpful for sparking creativity, as the ambiguity removed focus from specific details and allowed for more open idea generation. Many participants also found the partially denoised images sufficient for assessing the direction of their work, thus saving both time and energy. When explained, the concept of energy savings further shifted opinions positively on partial denoising, with participants more inclined to embrace it once they understood its environmental benefits. This suggests that partial denoising, when communicated effectively, could offer a balanced option that promotes creativity while supporting sustainability and efficiency.

Supporting the findings of Paananen et al., multiple participants mentioned how a more spatial layout might better facilitate the branching of ideas than the linear structure of models such as \textit{Midjourney} and \textit{DALL-E}. Compared to their study, however, our system received lower CSI scores across all conditions. In the study by Paananen et al.\ on architectural students, popular image generators achieved mean scores ranging from 70.9 to 75.2~\cite{paananen2023architecture}, whereas our system scored between 60.5 and 67.6\footnote{Paananen et al. included \textit{Cooperation} as an optional dimension for participants to rate. Our exclusion of this dimension makes direct comparisons difficult}. This disparity may partly stem from the prototype nature of our tool, which employs a simple model designed for testing specific conditions, unlike the commercial, fully developed tools evaluated in the study by Paananen et al. Subject matter differences could also be a factor; generating images of people is inherently different from creating architectural visuals, and such images may more easily evoke the uncanny valley effect. For example, the \textit{Juggernaut Reborn} model often struggled with rendering eyes and fingers convincingly, causing many generated characters to appear noticeably `off'. This might be the reason some participants preferred partially denoised images, and were disappointed when images were fully denoised.

We found a significant negative correlation between duration/energy consumption and CSI score, indicating that task completion time is important for creativity support. This finding is reinforced by interview statements from seven participants, who expressed a preference for certain versions because they achieved results that better matched their expectations within the five-minute time frame. This suggests that participants might have perceived the tool more favourably when they efficiently achieved their desired images.
While this makes sense in scenarios where the user already has an intended image in mind, this might not be beneficial for a tool intended to support in ideation. In this context, the objective is not merely to quickly obtain a specific image, but rather to explore diverse directions. While participants rated exploration as the most valued dimension, the preference for quick task completion suggests a potential tension between these goals. Balancing speed with opportunities for meaningful exploration is a noteworthy consideration in designing future AI tools. From the post-test questionnaire and interviews we learned that most participants did not consider their environmental impact when using AI tools, believing that sustainability is a corporate rather than a personal responsibility. Even then, many expressed interest in being informed about their consumption, particularly when the data was presented in relatable terms. While research indicates that this type of communication can increase awareness~\cite{ren2023guiltyinagoodway}, participants were unsure whether such eco-feedback would actually change their behaviour.

Compared to prior work by Luccioni et al.~\cite{luccioni2024powerWatts}, our setup demonstrated significantly lower energy consumption per generated image. Specifically, generating a fully denoised image in our study consumed approximately 0.3 Wh %(see Appendix~\ref{apx:energy_consumption})
, roughly ten times less than the 2.907 Wh per image reported by Luccioni et al. While our study exclusively used \textit{Juggernaut Reborn}, Luccioni et al. evaluated multiple image generators, including models based on Stable Diffusion. The higher energy costs reported in their work are likely due to the increased parameter counts, higher image resolutions, and greater number of steps used by some of these models. For example, Stable Diffusion XL is optimised for images that are 1024x1024 pixels~\cite{podell2023sdxl}, which is four times larger than the 512x512 used by \textit{Juggernaut Reborn}.

\subsection{Limitations and Future Work}
We imposed the five-minute time limit per condition to maintain the study's feasibility. However, this restriction might have impacted the depth of creative exploration as many participants ran out of time. 
Due to hardware limitations when developing the app we used a \textit{Juggernaut Reborn} image generation model based on Stable Diffusion 1.5, which is older, faster, and less demanding than new high performance models like DALL-E 3 and Midjourney.  
Finally, our specific model may not have the same energy usage characteristics as other popular models (less or more energy spent on upscaling compared to denoising steps) which may affect the generalisability of the findings. Our reliance on \textit{CodeCarbon}, which we chose for its simplicity and ease of implementation, may also represent a limitation making comparisons to future AI research more difficult. In this study, we selected an initial denoise level of 20 out of 30 steps, as we considered this the minimum threshold at which the contents of the image could be distinguished. The initial denoise level could be studied further to identify the optimal trade-off between image quality and energy usage. Likewise, the selected prompt-to-image ratio of 1:4 was based on mainstream image generation tools like \textit{Midjourney} and \textit{Adobe Firefly} and could be studied further. Many participants expressed a desire for the ability to make small, targeted changes rather than relying on generating entirely new images. Current workflows often require regenerating images with slightly adjusted prompts, which can result in very different outputs. Studying how inpainting could improve both creative control and energy consumption could therefore be a valuable next step. 
As AI video generation tools, such as OpenAI's Sora, become more prevalent, studying their energy consumption and creativity support during ideation will be increasingly relevant. Video generation demands significantly higher computational resources than image generation~\cite{li2024carbon}, making it even more important to explore sustainable approaches in this domain. Finally, eco-feedback systems, which provide users with real-time information about their energy consumption and environmental impact, represent a promising avenue for future exploration. While participants showed interest in such features, their actual influence on user behaviour in the long term is unclear. Future studies could implement eco-feedback mechanisms and measure their long-term effects on both energy usage and user interaction patterns.

\section{Conclusion}
This study explored the effects of partial denoising and batching (i.e., the quality and quantity) of generated images per prompt) in AI image generation, focusing on effects of energy consumption and creativity support. The findings confirmed that partial denoising and single-image generation consumed significantly less cumulated energy than full denoising and batching (four-image generation) throughout a complete design tasks. We found no significant differences in creativity support, perhaps suggesting that image generators can be designed to lower energy consumption without diminishing their ability to support creativity. Qualitative insights, however, revealed that a majority of participants preferred batching as it provided inspiration through diverse options. Similarly, many favoured full denoising for its clarity and smoother interaction, while others appreciated the partially denoised images for their ability to remove focus from specific details, fostering a more open and exploratory ideation process. The analysis of how users prompts changed through out the task provided insights into the system's impact on the creative process of converging and diverging to or from an idea. These results highlight the potential for configurable and adaptive settings, allowing a balance between sustainability and creative flexibility. We hope that fellow researchers will complement our research to drive progress in enabling sustainable forms of high performing creativity support. By aligning sustainable practices with effective design, this study contributes to the evolving discourse on Green AI in HCI.

\bibliographystyle{ACM-Reference-Format}
\bibliography{bibtex.bib}

\clearpage

\end{document}